\documentclass{PoS}

\title{Lattice QCD with 
Classical \\
and Quantum Electrodynamics}

\ShortTitle{Lattice with Electrodynamics}

\author{\speaker{B.~C.~Tiburzi}\\
        Center for Theoretical Physics\\
        Massachusetts Institute of Technology\\
        Cambridge, MA 02139, USA\\ \\
        Department of Physics\\ 
        The City College of New York\\  
        New York, NY 10031, USA\\ \\
        Graduate School and University Center\\
        The City University of New York\\  
        New York, NY 10016, USA\\  \\
        RIKEN BNL Research Center\\ 
        Brookhaven National Laboratory\\ 
        Upton, NY 11973, USA\\ \\
        E-mail: \email{btiburzi@ccny.cuny.edu}
        }

\abstract{
We are doubtlessly familiar with some edition of Jackson's tome on electrodynamics, 
and Schwinger's calculation of the anomalous magnetic moment of the electron in QED. 
From the perspective of strong interactions, 
however, 
electromagnetic effects usually amount to negligible contributions. 
Despite this fact, 
electromagnetic probes have always been a fundamental source for our knowledge of QCD experimentally. 
Elastic scattering of electrons off nucleons provides us a window to their distributions of charge and magnetism. 
To account for the spectrum of QCD at the percent level, 
moreover, 
we need isospin breaking introduced from both quark masses and electric charges. 
This overview concerns some of the prospects and progress of studying electromagnetic effects in QCD. 
Our focus is divided between classical and quantum effects. 
In classical electromagnetic fields, 
the dynamical response of QCD to external conditions can be investigated. 
The vacuum and hadrons alike should be viewed as media which respond to external fields: 
both magnetize and polarize in magnetic fields, for example. 
At the quantum level, 
electromagnetism and QCD renormalize each other. 
In the era of high precision lattice computations, 
both strong and electromagnetic contributions must be accounted for to make predictions at the percent level.
}

\FullConference{The XXIX International Symposium on Lattice Field Theory, Lattice2011\\
		July 10-16, 2011\\
		Squaw Valley, Lake Tahoe, California}

\begin{document}  

\section{Scope}

Electrodynamics in both its classical and quantum form is a staple in the study of physics. 
While the electromagnetic interactions of quarks are dwarfed by their strong interactions, 
electromagnetic observables and small electrodynamic effects can ultimately give one an intuitive 
picture of the underlying QCD environment present in the vacuum and in hadrons. 
In the era of high precision lattice QCD, 
moreover, 
precise quantitative understanding of the strong interaction requires accounting for electromagnetic effects. 
The purpose of this overview is to touch briefly upon a few topics involving QCD in the presence of electromagnetism. 
In the past few years, 
the level of activity in this field has grown considerably. 
Various different electromagnetic effects are currently being addressed using lattice QCD techniques; 
and, 
we expect many of these calculations to be refined in the foreseeable future.  We do not attempt a complete review of this subject. 
Our goal is to give one a general picture of the prospects and progress in the field by highlighting just a few areas. 
In many of these areas, 
a detailed technical review of the subject is not yet warranted. 
We look forward to the many future developments that will likely necessitate a thorough review of this field. 
Our overview is divided into two parts: 
one dealing with aspects of QCD in classical electromagnetic fields, 
and a second dealing with aspects of QCD in the presence of QED.

\section{Lattice QCD with Classical Electrodynamics}

Studying QCD in the presence of classical electromagnetic fields allows one to address how QCD responds to external conditions. 
We will focus on just two aspects of this response. 
First, we look at how hadrons respond to weak electric fields. 
Second, we consider how the vacuum of QCD responds to strong magnetic fields.

\subsection{Weak Electric Fields}

Materials are characterized by two broad types of properties. 
Materials have intrinsic properties that are determined by their underlying structure. 
On the other hand, 
materials also have extrinsic properties that are determined by their response to external conditions. 
For hadrons, 
very little is known about their extrinsic properties. 
The only extrinsic properties reported in the PDG concern the response of the nucleon to external electromagnetic fields. 
This response is captured by the electromagnetic polarizabilities.

Electric polarizabilities are familiar from classical electrodynamics. 
A material without an electric dipole moment will nevertheless polarize in an applied electric field. 
The induced dipole moment, 
$\vec{d}_E$, 
will be proportional to the applied field and oppositely oriented,
$\vec{d}_E = - \alpha_E \vec{E}$,
with a coefficient, 
$\alpha_E$,
representing the electric polarizability of the material. 
Consequently the effective Hamiltonian for the material in an external field acquires a term at second order in the electric field, 
$\Delta H = - \frac{1}{2} \alpha_E \vec{E} \, {}^2$, 
with higher-order terms assumed to be negligibly small. 
Electric polarizabilities of atoms and molecules play a crucial role in chemistry and biology. 
Quantum mechanically
one can compute the polarizabilitiy of the ground-state hydrogen atom exactly, 
$\alpha_E^{H-atom}
=
\frac{27}{8 \pi}
\left( \frac{4}{3} \pi a_B^3 \right)
\propto
e^{-6} \, m_e^{-3}$.
Not surprisingly the polarizability scales with the volume of the hydrogen atom ground state with a coefficient that is order unity. 
In the limit of vanishingly small electric charge, 
the hydrogen atom becomes extremely loosely bound, 
and the electron cloud is readily polarized by the applied field. 
Furthermore, 
in the limit of vanishingly small electron mass, 
the hydrogen atom becomes completely diffuse.

For hadrons, 
we thus have considerable intuition about their polarizabilities, 
the form of which must scale with the typical hadron's volume: 
$\alpha_E^{Hadron}
=
\mathcal{N}
\alpha_{f.s.}
\left( \frac{4}{3} \pi \, [\texttt{fm}^3] \right)
$,
with 
$\mathcal{N}$
a hadron-dependent pure number. 
In the limit of vanishingly small electric charge, 
the electric field decouples from the quarks. 
The scale of the hadron's volume, 
however, 
is a dynamically generated quantity depending on non-perturbative QCD.  
It is consequently less obvious how the electric polarizability should behave near the chiral limit, 
$m_q \to 0$. 
In the MIT bag model~\cite{Chodos:1974je}, 
the nucleon polarizability is insensitive to the quark mass, 
$\alpha_E \sim e^2 R_{bag}^3$ 
in the chiral limit; 
whereas, 
in chiral perturbation theory, 
the polarizability has a striking non-analytic behavior, 
$\alpha_E \sim e^2 \, m_q^{-1/2} \Lambda_{QCD}^{-1/2} \, \Lambda_{\chi}^{-2}$,
associated with the charged pion cloud~\cite{Bernard:1991rq}. 
In the chiral limit, 
hadrons should just barely become diffuse; 
but, 
whether the physical values of the light quark masses put us in the regime where this term dominates is unclear. 
For the charged pion, 
predictions from chiral perturbation theory for polarizabilities are almost a factor of two smaller than the most recent experimental determination~\cite{Holstein:1990qy}.

Electromagnetic polarizabilities are accessible in Compton scattering experiments, 
and there are on-going experimental efforts to determine these quantities precisely. 
The COMPASS experiment at CERN hopes to provide the final word on charged pion polarizabilities, 
and the first word on charged kaon polarizabilities. 
Low-energy Compton scattering on deuterium has been performed recently at MAX-Lab in Lund to allow for the 
extraction of neutron polarizabilities. 
A significant effort is underway using HI$\gamma$S at TUNL to measure Compton scattering off hydrogen and deuterium targets. 
It is thus timely for the lattice to compute electromagnetic polarizabilities to help resolve long-standing discrepancies. 
The ability to vary the quark masses is a virtue, 
moreover, 
as it allows for a crucial test of low-energy QCD from first principles.

The Compton scattering tensor between hadrons is a four-point function, 
namely
$T_{\mu \nu}(k,k') 
= 
\int_{x,y} e^{ i k \cdot x - i k' \cdot y} \langle H | \mathcal{T} \left\{ J_\mu (x) J_\nu(y) \right\} | H \rangle$, 
which lattice techniques have not yet tackled. 
Even if the full four-point function could be evaluated on the lattice, 
access to polarizabilities would only be through long extrapolations to vanishing photon momentum. 
Fortunately there exists an alternative that dates to the early period of quenched lattice QCD:
the background field method. 
Background magnetic fields were used in the first calculations of nucleon magnetic moments~\cite{Bernard:1982yu},
and shortly after it was realized that effects at second order in background electric fields could be used to extract 
electric polarizabilities of neutral hadrons~\cite{Fiebig:1988en}.

The background field method was revisited a few years ago to determine
magnetic moments and magnetic polarizabilities of hadrons, 
and electric polarizabilities of neutral hadrons~\cite{Lee:2005ds};
these studies were completely quenched.  
Non-uniform external fields were employed by these studies with effects from field gradients 
mitigated by imposing Dirichlet boundary conditions in space (space and time) 
for magnetic (electric) fields. 
The corresponding eigenstates are standing waves, which were not addressed in these works. 
The boundary conditions, 
moreover, 
introduce non-perturbative effects that are difficult to quantify.

Further computations have utilized external electric fields to deduce the 
electric dipole moment of the nucleon in the presence of the $\theta$-term~\cite{Shintani:2006xr}.
These computations are partially quenched 
(in this case, they are quenched only by shutting off the electric charges of the sea quarks), 
but still suffer a spike in the electric field at the temporal boundary.
Here standard antiperiodic boundary conditions are imposed on the quarks in time, 
with sources and sinks located as far as possible from the spike.   
Additional studies have given preliminary reports on the pion mass dependence and volume dependence of 
neutral hadron electric polarizabilities, 
and a comparison of boundary conditions%
~\cite{Alexandru:2009id}.\footnote%
{
It is of considerable interest to note an alternate approach that is complimentary to the background field method.
Instead of including background fields in the action and later hunting for effects in two-point functions that are a desired order in the strength of the external field, 
one could expand correlation functions from the outset to isolate physics at the desired order. 
This alternate approach has been developed and demonstrated by measuring the electric polarizability of the neutron
including all disconnected terms for $m_\pi = 760 \, \texttt{MeV}$~\cite{Engelhardt:2007ub}.
A salient feature of this approach is the ability to treat explicitly volume effects from the holonomy of the external field. 
A downside is that non-uniform fields are used to perturbatively expand correlators. 
}
We summarize three recent developments in external electric field computations.

\begin{figure}
\begin{center}
\includegraphics[width=6cm]{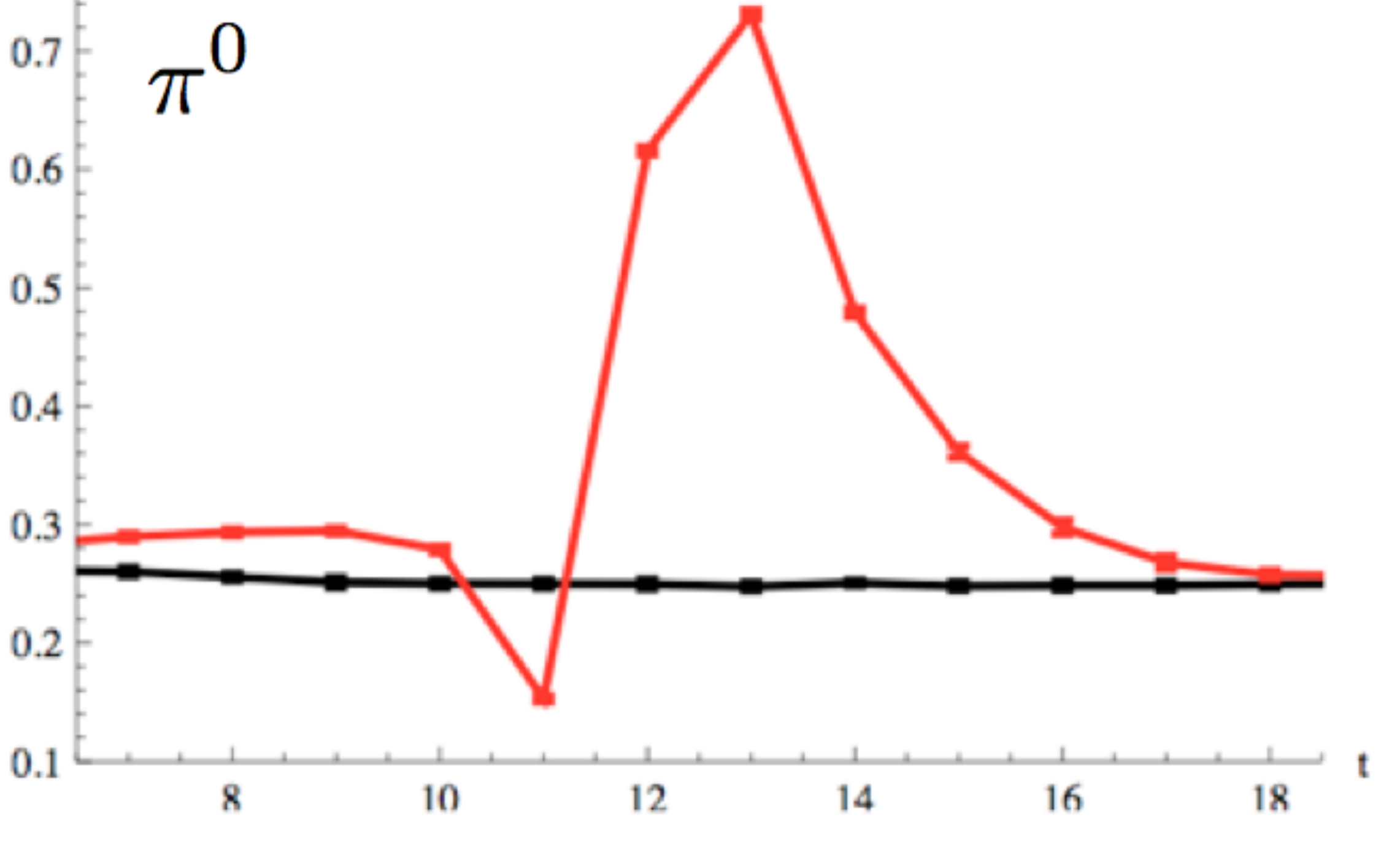}
\includegraphics[width=6cm]{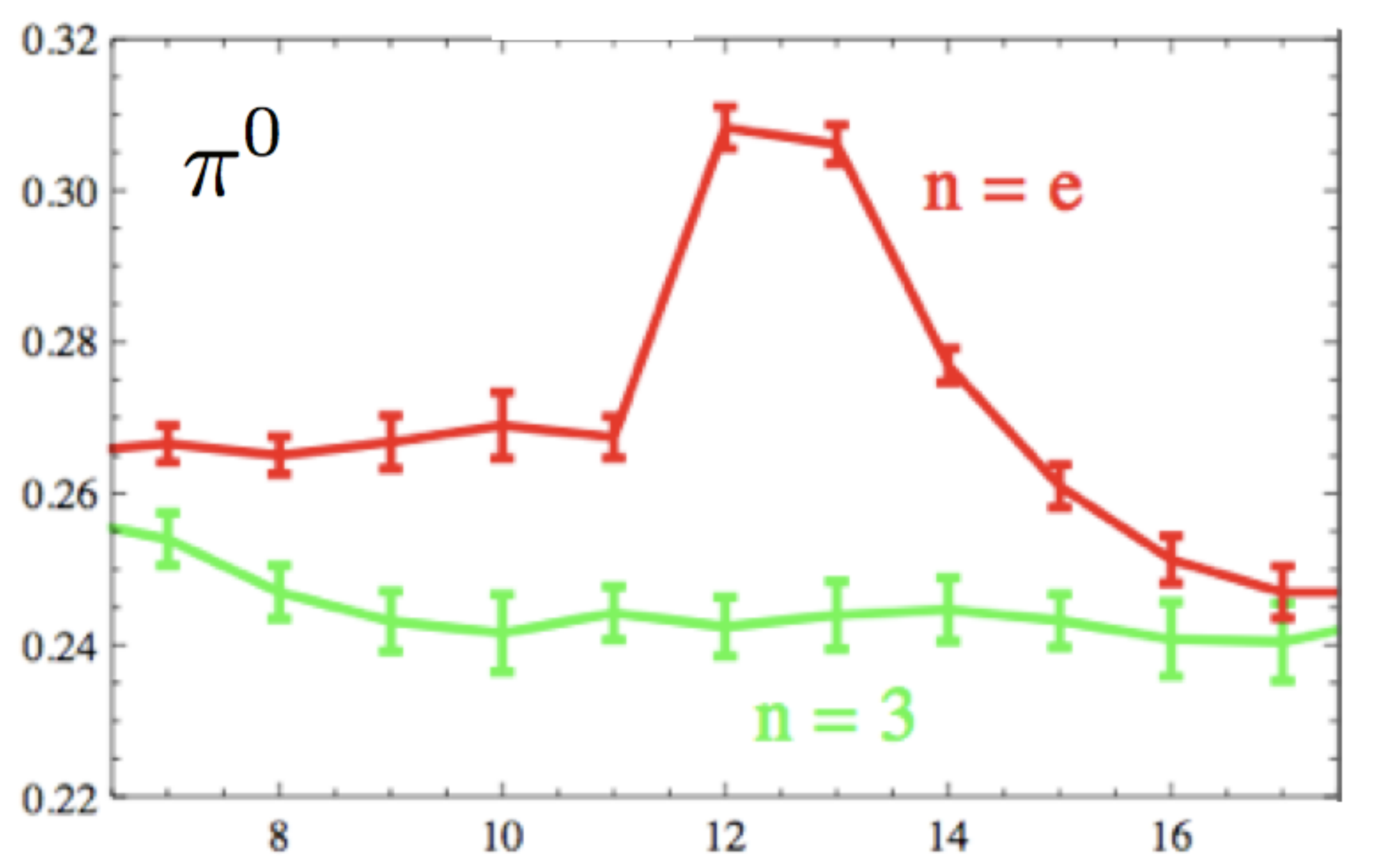}
\end{center}
\caption{\label{f:spike} 
Effective mass plots for (connected) neutral pion correlation functions in external fields.
On the left, 
we compare zero field and the na\"ive implementation of an electric field. 
On the right, 
we compare quantized, 
$n = 3$,  
and non-quantized, 
$n = e = 2.71828\ldots$, 
values of the electric field with a boundary twist included. 
In both cases, 
the correlators have been translated in time so that 
$t=12$ 
corresponds to the edge of the lattice. 
Plots are taken from~\cite{Detmold:2008xk}.
}
\end{figure}
%
%

{\bf 1).}~To apply a uniform electric or magnetic field on a torus, 
quantization conditions must be met%
~\cite{'tHooft:1979uj}. 
The generalization for a discrete torus, 
i.e.~a lattice, 
has been detailed in%
~\cite{Smit:1986fn}.
For example, 
the allowed strengths of uniform 
(Euclidean) 
electric fields
$\mathcal{E}$
satisfy
\begin{equation} \label{eq:quant}
q \mathcal{E} = \frac{2 \pi n}{ \beta L}
,\end{equation}
with 
$q$ 
as the quark electric charge, 
and 
$n$ 
an integer.\footnote%
{
While quantized field strengths eliminate the gradient systematic, 
there are still non-trivial finite size effects.
These arise due to non-vanishing Polyakov loops, 
e.g.  the na\"ive electric field has the holonomy  
$\Phi_3 = \int_0^L dx_3 A_3(x) = - \mathcal{E} L x_4$. 
At the level of hadrons, 
charged pions propagating around the world $n$-times must be dressed with a (spatial) Polyakov loop, 
$e^{ i n \Phi_3}$. 
Finite volume effects thus introduce a time-dependent potential, 
however, 
coefficients in this potential are exponentially suppressed, 
$\propto e^{- n \, m_\pi L}$
for asymptotically large volumes, 
see~\cite{Tiburzi:2008pa}. 
} 
Lattices are currently large enough to support perturbatively small electric fields. 
Effects of quantized versus non-quantized fields were studied in%
~\cite{Detmold:2008xk}.
For a neutral pion in a uniform electric field, 
the energy has the form,
$E_{\pi^0} = m_{\pi^0} + \frac{1}{2} \alpha_E^{\pi^0} \mathcal{E}^2$, 
up to higher-order corrections in the field strength.  
With a non-uniform field, 
there are a tower of additional terms involving gradients. 
Each term in the tower is accompanied by an unknown coefficient.  
The electric field 
$\mathcal{E} \hat{z}$, 
can be specified by a time-dependent vector potential 
$A_\mu = ( 0, 0, - \mathcal{E} x_4, 0)$.  
To include this field on the lattice, 
one uses Abelian gauge links, 
$U_\mu = \exp ( i q A_\mu )$.
This na\"ive implementation leads to a large field gradient between 
$x_4 = \beta - 1$ and $x_4 =0$. 
A very dramatic effect can be seen on the (connected) neutral pion correlation function, 
as shown in Figure~\ref{f:spike}. 
The field gradient can be mitigated by including a boundary twist, 
$U_\mu^{twist} = \exp ( i q A_\mu^{twist} )$, 
with 
$A_\mu^{twist} = (0,0,0, \beta \mathcal{E}  x_3 \delta_{x_4, \beta -1} )$.  
The resulting field is uniform except for a huge gradient at the corner of the lattice, 
$(x_3, x_4) = (L-1, \beta -1)$. 
This gradient can be eliminated when the field is quantized as in Eq.~(\ref{eq:quant}), 
see Figure~\ref{f:spike}.

\begin{figure}
\begin{center}
\includegraphics[width=8cm]{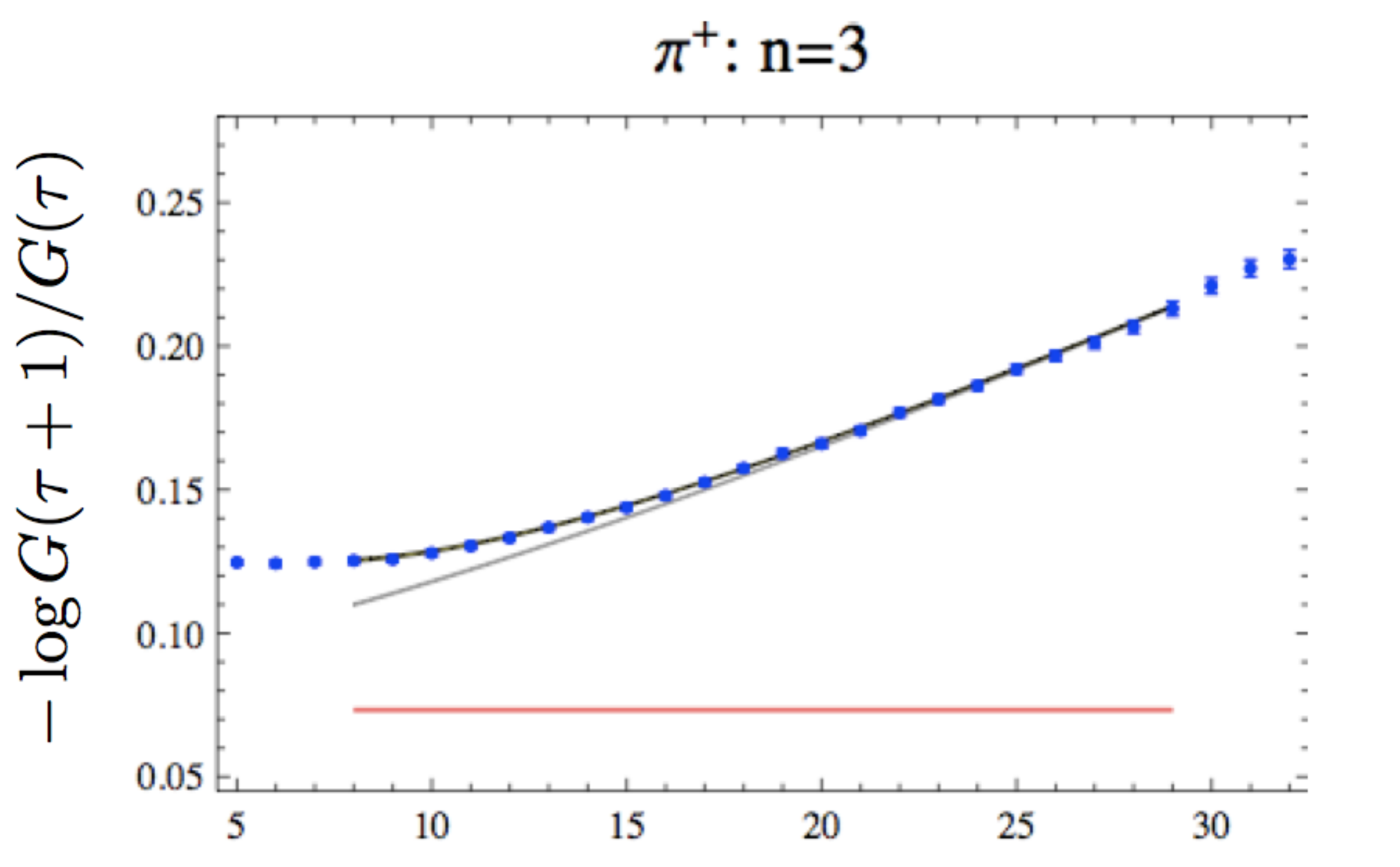}
\end{center}
\caption{\label{f:accel} 
Effective mass plot for the charged pion correlator calculated in an external electric field, 
with strength  
$n = 3$.
A fit to the correlator is shown allowing for excited-state contamination, 
and the extracted ``rest energy'' is plotted as the solid red line.
Plot taken from~\cite{Detmold:2009dx}.
}
\end{figure}

{\bf 2).}~Charged particle properties can be deduced in external electric fields. 
This was first suggested in%
~\cite{Detmold:2006vu}, 
where a non-relativistic method was developed. 
In light of the size of the external field strengths,  
a relativistic generalization is needed in practice,
and can be found in%
~\cite{Tiburzi:2008ma}. 
The basic idea behind the treatment of charged particle correlators, 
is to write down an effective action for the hadron in the external field.\footnote%
{
There is also implicitly the assumption that the spectrum of lowest-lying states remains ordered as in zero field. 
This assumption can break down in considerably strong fields, 
e.g.~%
$\mathcal{E}  \sim M_*^2$, 
where 
$M_*$ 
is an excited state mass. 
} 
For example, 
a composite scalar 
$\phi$
of charge 
$Q$ 
in an external electric field,
$A_\mu = ( 0, 0, - \mathcal{E} x_4, 0)$, 
has an effective action,
\begin{equation} \label{eq:effA}
\mathcal{L}_{eff} 
\left(
\vec{p} = 0, x_4
\right) 
= 
\phi^\dagger(x_4) 
\left[ 
- \frac{\partial^2}{\partial x_4^2} 
+ 
Q^2 \mathcal{E}^2 x_4^2 
+ 
E(\mathcal{E})^2  
\right] \phi(x_4),
\end{equation}
where the ``rest energy,''
$E(\mathcal{E}) = m_\phi + \frac{1}{2} \alpha_E^\phi \mathcal{E}^2 + \cdots,$
includes non-minimal couplings such as the polarizability. 
The propagator for Eq.~(\ref{eq:effA}) can be expressed easily using 
Schwinger's proper-time trick, 
\begin{equation} \label{eq:schwing}
G(\tau)
= 
\langle \tau | \frac{1}{2 \mathcal{H} + E^2} | 0 \rangle
=
\frac{1}{2}
\int_0^\infty
ds \, e^{ - \frac{1}{2} s E^2}
\langle \tau | e^{ - s \mathcal{H}} | 0 \rangle
.\end{equation}
Charged pions and kaons were investigated in external electric fields%
~\cite{Detmold:2009dx}, 
and their two-point correlators were fit using Eq.~(\ref{eq:schwing}).
Extracted values for $E(\mathcal{E})$ at various field strengths give one
the ability to determine charged particle electric polarizabilities. 
An effective mass plot for a charged pion correlation function is shown in 
Figure~\ref{f:accel}.

%
%

{\bf 3).}
Spin-$\frac{1}{2}$ particles have also been treated in electric fields. 
There is a subtlety even for the neutron because of the spin interaction with the external field. 
Using the neutron effective action, 
$\mathcal{L}_{eff} = N^\dagger \left[ \gamma_\mu \partial_\mu + E(\mathcal{E}) - \frac{\mu}{4 M} \sigma_{\mu \nu} F_{\mu \nu} \right] N$,
one finds that unpolarized neutron two-point functions are given by the standard exponential,
$Z \exp ( - E_{eff} \tau)$,  
with a falloff determined by 
$E_{eff} = M + \frac{1}{2} \mathcal{E}^2 \left( \alpha_E - \frac{\mu^2}{4 M^3} \right)$. 
The additional term is a 
$\sim 10 \%$ 
effect at the physical pion mass, 
but we should make no assumption of the effect at larger-than-physical values of the pion mass. 
If we can find a way to remove this term, 
moreover, 
we will have succeeded in determining both the electric polarizability and magnetic moment of the neutron. 
This problem was discovered in%
~\cite{Detmold:2010ts}, 
where a solution was posed and verified using lattice QCD. 
By analogy with a magnetic field, 
where 
$\frac{1}{2} \sigma_{\mu \nu} F_{\mu \nu} = \vec{S} \cdot \vec{B}$
and one forms spin-projected correlation functions to access magnetic moments, 
in an electric field,
we have
$\frac{1}{2} \sigma_{\mu \nu} F_{\mu \nu} = \vec{K} \cdot \vec{E}$, 
where the matrices 
$\vec{K}$
are the 
spin-$\frac{1}{2}$ 
generators of boosts. 
In forming boost-projected correlators, 
i.e.~by tracing with 
$\frac{1}{2} ( 1 \pm K_3)$, 
one finds
$G_\pm(\tau) = Z \left( 1 \pm \frac{\mu \mathcal{E}}{2 M^2} \right) \exp ( - E_{eff} \tau)$. 
Thus a simultaneous fit to these correlation functions will lead to a determination of 
$\mu$ 
and 
$\alpha_E$. 
The additional magnetic moment squared term leads to a $50 \%$ effect at 
$m_\pi = 390 \, \texttt{MeV}$. 
A generalization for charged spin-$\frac{1}{2}$ particles was also demonstrated in%
~\cite{Detmold:2010ts}.

{\bf At this conference:}
There were a number of developments related to the study of hadrons in weak external fields reported. 
A preliminary look at quenched neutron correlation functions in uniform magnetic fields was given%
~\cite{Primer}.
Because the charge and anomalous magnetic moment interactions do not commute for relativistic spin-$\frac{1}{2}$ particles, 
Landau levels must be treated to extract the magnetic moments and polarizabilities. 
As chiral symmetry properties of lattice fermions may play an important role in polarizability calculations,
issues related to using overlap fermions in external fields were discussed%
~\cite{Lujan}.
When one considers external fields with slowly varying space-time dependence, 
additional quantities, 
known as spin-polarizabilities, 
characterize the hadron's response. 
Such quantities are even more tightly constrained in chiral perturbation theory. 
Feasibility of extracting spin polarizabilities using background fields was discussed%
~\cite{Lee}.
A promising alternate method to compute the electric spin polarizability of the neutron was unveiled%
~\cite{Engelhardt}.

\subsection{Strong Magnetic Fields}

\begin{figure}[b]
\begin{center}
\includegraphics[width=8cm]{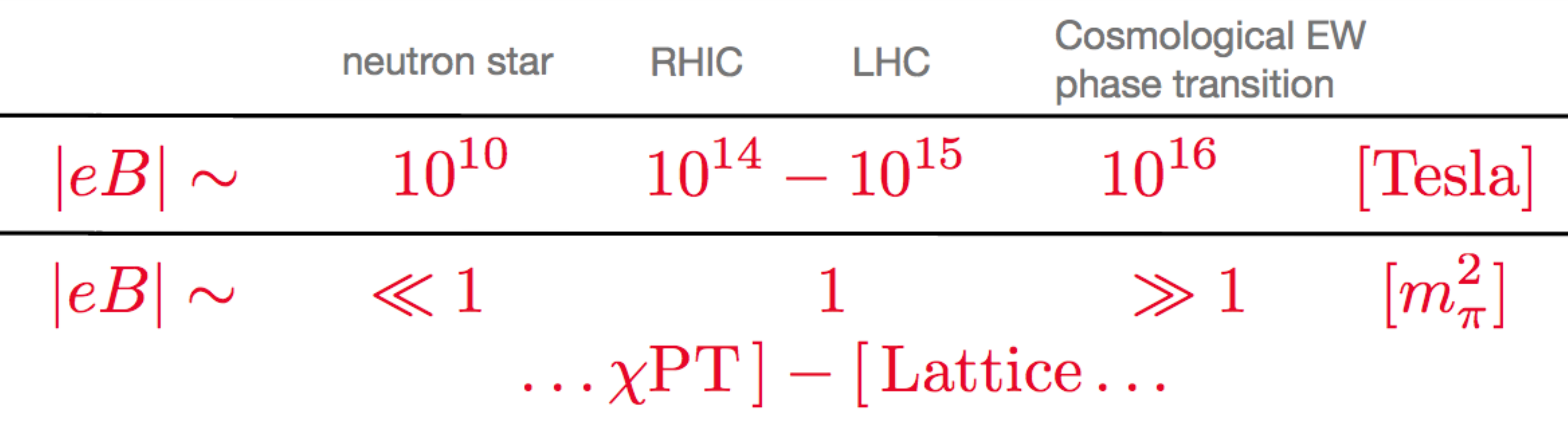}
\end{center}
\caption{\label{f:strong} 
Estimates of strong magnetic fields in the universe given in units of
$\texttt{Tesla}$, 
and 
$m_\pi^2$.
}
\end{figure}
%
%

Lattices are not yet large enough to support several values of perturbatively small, 
uniform magnetic fields. 
This is because the 't Hooft quantization condition for magnetic fields gives values
$q B = 2 \pi n / L^2$,
which are larger than the allowed electric fields in Eq.~(\ref{eq:quant}), 
because generally one has
$\beta \approx 2 L$. 
The dimensionful parameter determining whether a field strength is perturbative
is the mass of the lightest charged state that couples to the hadron in question. 
This state must be the pion due to QCD inequalities. 
Whether the pion is the dominant contribution, 
as in chiral perturbation theory, 
is irrelevant. 
Thus the expansion is governed by 
$e^2 B^2 / m_\pi^4 \sim n^2 / 7$,
for $m_\pi L \approx 4$, 
and one is certainly restricted to $n < 3$.  
Notice that 
$e^2 \mathcal{E}^2 / m_\pi^4$ 
is smaller by a factor of 
$\beta^2 / L^2 \approx 4$.  
Using uniform magnetic fields, 
there has been initial work on the magnetic moments of 
spin-$\frac{3}{2}$ 
baryons~\cite{Aubin:2008qp}, 
and an investigation of the magnetic-field dependence of the proton's wavefunction%
~\cite{Roberts:2010cz}.
Complete studies in weak, uniform magnetic fields await larger lattices.

\begin{figure}[b]
\begin{center}
\includegraphics[width=8cm]{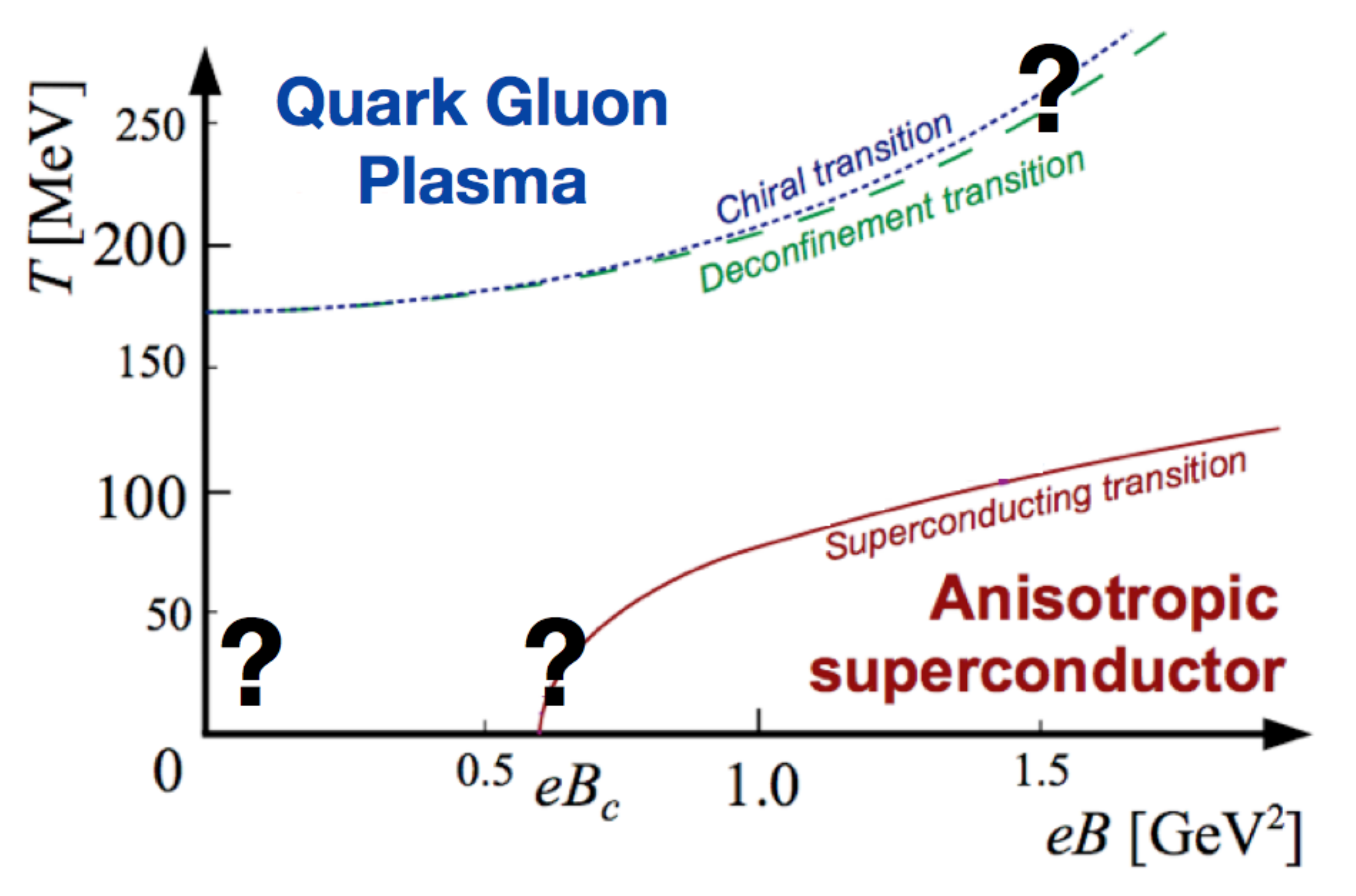}
\end{center}
\caption{\label{f:phases} 
A few elements of the conjectured phase diagram of QCD in the $B$--$T$ plane. 
We focus on questions in regions marked by question marks. 
Plot adapted from~\cite{Chernodub:2011tv}, 
where further details can be found. 
}
\end{figure}

Rather interesting effects can arise in strong magnetic fields;
and, 
these fields are not only accessible to lattices today, 
they are in a range relevant for heavy ion collisions. 
The currents generated by two beams of large ions
(be they gold or lead)
give rise to tremendous magnetic fields in the region 
between the two colliding beams. 
Estimated physical extremes for magnetic fields are summarized in Figure~\ref{f:strong}. 
Some elements of the conjectured phase diagram for QCD in an external magnetic field are 
depicted in Figure~\ref{f:phases}. 
We focus on three regions in the 
$B$--$T$ 
plane.

%
%

{\bf 1).} 
For 
$|e B| / m_\pi^2 \lesssim 1$
and
$|e B| / \Lambda_\chi^2 \ll 1$, 
one can use chiral perturbation theory with strong-field power counting to determine how QCD responds. 
For simplicity, 
we consider cold QCD, 
i.e.~temperatures such that 
$T \ll m_\pi$. 
It is known that a magnetic field catalyzes chiral symmetry breaking:
a magnetic field preserves configurations with particle-antiparticle pairs of opposite momenta. 
In the chiral limit, 
one has the non-analytic enhancement factor,
$\langle \overline{\psi} \psi \rangle_B  / \langle \overline{\psi} \psi \rangle =  1 + \log 2 |eB| / \Lambda_\chi^2 + \mathcal{O}
\left( \frac{m_\pi^4}{|e B|^2} \right)$,
calculable in chiral perturbation theory~\cite{Shushpanov:1997sf}. 
Low-energy theorems, 
such as the Gell--Mann-Oakes-Renner relation
can be generalized in external fields. 
A quantity of particular interest is the magnetization, 
$\chi$,
of the vacuum, 
which we define by, 
$\chi \equiv \lim_{B \to 0} \, \chi_B$, 
where
$\langle \overline{\psi} \sigma_{\mu \nu} \psi \rangle_B = \chi_B \, F_{\mu \nu}$. 
This parameter is of interest for QCD sum-rule calculations, 
and can be accessed experimentally by measuring chiral-odd parton distributions. 
Unfortunately it is an unknown parameter in chiral perturbation theory.\footnote%
{
One can use chiral perturbation theory to demonstrate that
$\chi_B / \chi = \langle \overline{\psi} \psi \rangle_B / \langle \overline{\psi} \psi \rangle$.
The constant of proportionality, 
however, 
is the parameter of interest. 
} 
Nonetheless, 
the lattice can fully explore the QCD response to strong magnetic fields in this regime and beyond.  
A pioneering lattice study has measured the response of the chiral condensate in quenched 
$SU(2)$ 
gauge theory%
~\cite{Buividovich:2008wf}, 
and later in 
quenched QCD%
~\cite{Braguta:2010ej}.
A further study gave the first treatment in two-flavor QCD%
~\cite{D'Elia:2011zu}, 
where fully dynamical lattice were generated with staggered quarks coupled to a uniform magnetic field.
Unquenching the magnetic field was found to have a substantial effect on the behavior of
the chiral condensate. 
Finally the magnetization of the vacuum was computed in quenched 
$SU(2)$
gauge theory%
~\cite{Buividovich:2009ih},
and quenched QCD%
~\cite{Braguta:2010ej}
using a generalization of the Banks-Casher relation.

{\bf 2).}
A second region of interest in the phase diagram encompasses the transition to the quark-gluon plasma. 
The chiral condensate melts as the temperature is raised, with chiral symmetry eventually restored above
some temperature. 
There is another finite temperature transition from a confined hadronic phase to a deconfined plasma phase. 
In zero magnetic field, 
these transitions occur at basically the same temperature. 
Turning on a magnetic field, 
however, 
it is conceivable that these transitions might substantially split%
~\cite{Mizher:2010zb}.
Na\"ively chiral symmetry breaking is enhanced in an external magnetic field; 
consequently we might expect chiral symmetry restoration to be postponed.
If this were the case, 
there could be a new phase of deconfined, chirally asymmetric matter.   
This possibility was first explored using lattice QCD in%
~\cite{D'Elia:2010nq}.
To find the deconfinement transition, 
typically the Polyakov loop is studied. 
Thus to assess the effect of a magnetic field on deconfinement, 
it is a necessity to have fully dynamical configurations. 
As mentioned above, 
these configurations were generated with a two-flavor staggered quark action. 
This study found a slight increase in 
$T_c$, 
but no noticeable split in the transitions. 
The transitions appeared to sharpen potentially becoming first order. 
Only one lattice spacing was employed, 
however, 
which was rather coarse,
$a \approx 0.3 \, \texttt{fm}$.  
Preliminary results from a more complete study were reported at this conference, 
see below.

{\bf 3).}
Our final look at strong magnetic fields concerns the possibility for novel QCD phase transitions.
Quenched lattice simulations show evidence for an insulator--(anisotropic) conductor transition in strong magnetic fields at non-zero temperatures%
~\cite{Buividovich:2010tn}.
As the temperature is lowered, 
there exists the possibility to form an anisotropic superconductor%
~\cite{Chernodub:2010qx}.
This is the QCD analogue of the electroweak vacuum instability in strong magnetic fields that leads to condensation of weak-bosons%
~\cite{Ambjorn:1988tm}.
In ordinary superconductors, 
strong magnetic fields will cause the Meissner effect to break down. 
For the present case,
however, 
past a critical value of the magnetic field, 
$B_c$, 
charged vectors should condense leading to a superconducting state, 
see Figure~\ref{f:phases}.

Unlike the point-like vector bosons of the weak interaction, 
vectors in QCD are composite particles with non-minimal couplings to the external field. 
The effective action for a charged rho-meson in a uniform external field is
\begin{equation}
\mathcal{L}_{eff}
=
\rho^\dagger_\alpha
\left[
(-D^2 + M_\rho^2) 
\delta_{\alpha \beta}
+ 
D_\beta D_\alpha
- 
i \mu_\rho F_{\alpha \beta}
\right]
\rho_\beta
.\end{equation}
In the action, 
we have included an anomalous magnetic moment coupling,  
$\mu_\rho$,
but ignored the field-strength dependence of the mass, 
$M_\rho$, 
and omitted the width. 
A charged pion must be one of the decay products of a charged rho, 
but the decay becomes suppressed in large magnetic fields because the 
lowest Landau level for the charged pion has an energy that increases with increasing field strength, 
$E_\pi = \sqrt{m_\pi^2 + |e B|}$.
The lowest Landau level for the charged rho, 
by contrast, 
has the form,
$E_\rho = \sqrt{M_\rho^2 -  \mu_\rho B }$,
when the rho is polarized opposite the direction of the magnetic field. 
In this simplified picture, 
the vacuum will become unstable past 
$B_c = M^2_\rho / \mu_\rho$. 
In this regime, 
a polarized charged rho condensate will form:
$\langle 0 | \rho^\dagger_{\downarrow} | 0 \rangle = \mathcal{F}(\vec{x} \times \vec{B})$, 
where 
$\mathcal{F}$ 
is a particular periodic function of the coordinates transverse to the magnetic field%
~\cite{Chernodub:2011mc}.
Using quenched 
$SU(2)$
gauge theory, 
quark condensates with charged-rho 
quantum numbers were found past a critical field 
$|e B_c| \sim \texttt{GeV}^2$%
~\cite{Braguta:2011hq}, 
which suggests that this picture might be correct for QCD.

{\bf At this conference:}
Results related to the phase structure of QCD in strong magnetic backgrounds were discussed.  
A brief summary was included in the review of QCD at finite temperature and density%
~\cite{Levkova}. 
The first study of the chiral condensate in QCD with a magnetic field 
(described above) 
was reviewed%
~\cite{Negro}.
A preliminary report of an extensive study of QCD in external magnetic fields was given%
~\cite{Endrodi}.
In this study, 
staggered lattices coupled to magnetic fields were generated with 
$2+1$ 
flavors at the physical quark masses. 
Ensembles with different lattice spacings were generated, 
moreover, 
to enable taking the continuum limit. 
There was no splitting of deconfinement and chiral transitions observed
(up to $|e B| \sim \texttt{GeV}^2$). 
Unlike the first study, 
however,
the critical temperature was found to decrease, 
with the transitions remaining crossovers.  
Data from this study show that magnetic catalysis of chiral symmetry breaking is not universal: 
there is an interplay between magnetic field, temperature, and quark mass dependence of the chiral condensate.

Relating to the chiral magnetic effect%
~\cite{Kharzeev:2007jp}, 
there were two talks. 
In~\cite{Ishikawa}, the charge separation of low modes was investigated in a magnetic field; 
while, in~\cite{Yamamoto:2011qa}, 
a so-called chiral chemical potential was employed.  
Previous lattice results on the effect include%
~\cite{Buividovich:2009wi,Braguta:2010ej}.
We regret that time and now space do not permit us a closer look.

\section{Lattice QCD with Quantum Electrodynamics}

In nature, 
strong and electromagnetic interactions coexist. 
Consequently they renormalize each other.
High precision work in lattice QCD requires treatment of electromagnetic effects.  
This has lead to the study of QED corrections to hadron properties.\footnote%
{
There are also cases for which QCD corrections are needed for QED observables. 
The most noteworthy example being the high-precision experimental determination of the anomalous magnetic moment of the muon. 
Confrontation of this result with theoretical calculations has been limited by our knowledge of the QCD corrections. 
The status of these non-perturbative corrections to QED computed in lattice QCD was reviewed at this conference~\cite{Renner}.
}
These studies are reviewed after we first describe a few QCD applications that require QED.

{\bf 1).}
Accounting for isospin splittings in the spectrum of QCD 
is one important problem that will be solved by treating QED effects on the lattice. 
In QCD + QED, 
$SU(2)_V$ 
is broken by the light quark mass difference, 
and quark electric charges.
In the nucleon mass, 
the leading isospin breaking 
is proportional to
$m_d - m_u$, 
and 
$q_u^2 - q_d^2$. 
The well-measured nucleon splitting is only a fraction of the nucleon mass,
$(M_n - M_p)  / M_N \sim + 0.1 \%$. 
The electromagnetic mass of the proton can be estimated crudely\footnote%
{
While this estimate is crude, 
it is consistent with chiral symmetry constraints. 
Treating the quark charges as arbitrary, 
the classical sphere estimate has the dependence,
$(M_n - M_p)_{QED} \propto Q_n^2 - Q_p^2 =  - 3 ( q_u^2 - q_d^2)$.

} 
 from the classical energy needed to build up a sphere of charge 
$|e|$
of radius 
$1 \, \texttt{fm}$,
which is 
$\propto \alpha_{f.s.} /  1 [\texttt{fm}] \sim + 1 \, \texttt{MeV}$.
Ignoring magnetic effects,  
the electromagnetic splitting is on the order of 
$(M_n - M_p)_{QED} \sim - 1 \, \texttt{MeV}$, 
from which we infer the strong isospin splitting, 
$(M_n - M_p)_{QCD} \sim + 2 \, \texttt{MeV}$. 
To determine the nucleon electromagnetic mass rigorously,
one needs to account for the distribution of both charge and magnetism within the nucleon. 
This information is provided by the elastic electromagnetic form factors. 
There are, 
however, 
additional inelastic contributions, 
and their theoretical status has been discussed recently, 
see%
~\cite{WalkerLoud:2010qq}
and references therein. 
The electromagnetic mass splitting can be computed using lattice QCD with QED.

Our discussion of isospin splitting in the nucleon brings up an important point. 
Currently our knowledge of the strong isospin splitting is limited by the removal of the electromagnetic contribution. 
This is a general feature. 
High precision determination of strong interaction parameters, 
most importantly: the masses of up and down quarks, 
is limited by our understanding of QED corrections to the inputs. 
The mass of the up quark is particularly susceptible to uncertainty arising from QED corrections to observables because 
it is the lightest quark.\footnote%
{
Na\"ively electromagnetic radiation involving the up-quark alone is $4 \times$ more important than that of the down-quark. 
Accounting for photon radiation at the quark level, 
however, 
is more involved, 
see%
~\cite{Goldman:1989as}, 
where some surprising implications for isospin splittings and Dashen's theorem are explored.
} 
One way to determine strong interaction parameters without additional assumptions is to simulate
$1+1+1$--flavor QCD + QED.

{\bf 2).}
Within the pseudoscalar meson sector, 
electromagnetic corrections to meson masses are universal in the chiral limit. 
This is known as Dashen's theorem, 
and it is thus interesting to measure its violation.
Violation of the theorem can be parameterized by a quantity
$\Delta M^2_{Dashen}$, 
where 
$\Delta M^2_{Dashen} = (m_{K^+}^2 - m_{K^0}^2)_{QED} - (m_{\pi^+}^2 - m_{\pi^0}^2)_{QED}$. 
By construction, 
$\mathcal{O}(\alpha_{f.s.})$
terms cancel out of  
$\Delta M^2_{Dashen}$, 
and the leading non-vanishing term has the behavior
$\mathcal{O}( m_s \, \alpha_{f.s.})$
away from the 
$SU(3)$ 
chiral limit. 
A comparison of lattice and phenomenological results for this quantity was shown in%
~\cite{Portelli:2010yn}, 
with the agreement decidedly poor.

{\bf 3).}
Beyond masses and mass splittings, 
isospin breaking effects can complicate the comparison of other strong interaction observables computed on the lattice to experiment. 
High precision determination of the kaon decay constant,
$f_K$,
is a good example of such a quantity.  
Experimentally the combination
$\frac{f_K}{f_\pi} \times | V_{us} | / |V_{ud}|$
is determined to an accuracy of 
$0.2 \%$. 
Lattice results on 
$f_K / f_\pi$
are known to 
$0.4 \%$, 
using the world average, 
and are used to extract the CKM matrix elements. 
At this level of precision, 
isospin breaking effects need to be considered. 
The experiment measures the charged kaon decay width, 
thus the relevant quantity is 
$f_{K^\pm}$.
There are electromagnetic corrections to the decay constants of 
$\mathcal{O}(\alpha_{f.s.})$ 
which have been accounted for in the analysis used to extract CKM matrix elements. 
The up-quark mass dependence was an additional issue considered in~\cite{Aubin:2004fs}.
Isospin degenerate simulations extract 
$f_{K}(\overline{m})$,
where 
$\overline{m} = \frac{1}{2} ( m_u + m_d)$
and the functional dependence indicates only the lightest valence quark mass, 
whereas one needs 
$f_{K}(m_u)$
for the charged kaon. 
Using chiral perturbation theory to account for the valence quark mass dependence leads to a 
$\sim 0.2\%$
difference in 
$f_{K^\pm}$.
Given the precision of the world average, 
it is of crucial importance to have a high precision determination of 
$m_u$
to account for such effects.
One also needs simulations at the physical value of 
$m_u$ 
to end the reliance on chiral perturbation theory.

\subsection*{Simulations of QCD + QED}

Methods to treat QCD + QED were pioneered some time ago%
~\cite{Duncan:1996xy}.
The total lattice action is now a sum of three terms:
$S = S_{YM} + S_{fermion} + S_{U(1)}$. 
Gauge invariance requires the 
$U(1)$
field to be coupled to fermions in 
$S_{fermion}$
through Abelian links, 
$U_\mu = \exp ( i q A_\mu)$. 
Gauge invariance of the photon action
$S_{U(1)}$
permits either compact or non-compact photon fields. 
A non-compact formulation of the 
$U(1)$ 
gauge field removes photon-photon interactions, 
so that the theory described by 
$S_{U(1)}$ 
alone is a free theory.
As a consequence, 
photon gauge configurations can be easily generated. 
Non-compact photon fields, 
however,  
require gauge fixing.

Notorious challenges of including QED interactions on a lattice are finite size effects.
As electric charges are not confined, 
the resulting long-range interactions suffer from being on a torus. 
For classical electromagnetic fields, 
field quantization emerges due to the intrinsic properties of the torus, 
namely the flux into any area 
$A$
in the $(i$-$j)$--plane must be the flux out of the area 
$L_i L_j - A$,
because the torus forms a closed surface.
Quantized field strengths still lead to finite-size artifacts due to the holonomy of the gauge field. 
These issues with classical electromagnetic fields on a torus are mirrored by related issues with their quantum counterparts. 
Because a torus can leak no flux, 
Gauss's law on the torus becomes trivial. 
The charge enclosed in the entire lattice volume vanishes by periodicity,
$Q = \int_0^L  J_4  \, d \vec{x} = \int_0^L \vec{\nabla} \cdot \vec{\mathcal{E}}  \, d\vec{x} = 0$.   
In simulations, 
this rather extreme finite volume effect is mitigated by deleting certain zero modes from the photon field. 
This non-local procedure amounts to forcing the longest-range components of the photon to be non-dynamical.
We summarize approaches taken so far by groups studying the combined QCD + QED dynamics.  

{\bf 1).}
Since the pioneering work of%
~\cite{Duncan:1996xy},
the first serious attack on the computation of QED corrections to QCD has been on domain-wall fermion lattices generated by the RBC collaboration%
~\cite{Blum:2007cy}.
In these studies, 
non-compact QED fields were generated, 
and domain-wall propagators inverted on post-multiplied gauge field configurations, 
i.e.~quenched QED.
The QED fields were fixed to Feynman gauge with spatial zero modes removed, 
$A_\mu(\vec{p} = 0, x_4) \equiv 0$. 
The physical electric charge 
$|e_{phys}|$
was employed, 
and a reduction of noise was achieved by averaging over positive and negative values of the charge, 
$e = \pm e_{phys}$.  
In the most recent study, 
ensembles with $2+1$--flavors have been used, 
with pion masses ranging from 
$250$--$700 \, \texttt{MeV}$. 
Results were obtained at a single lattice spacing, 
$a = 0.11 \, \texttt{fm}$, 
and on two volumes 
$V = (1.8 \, \texttt{fm} )^3$ 
and 
$(2.7 \, \texttt{fm} )^3$.  
Finite volume, partially quenched chiral perturbation theory + QED 
was used to account for  systematics.

{\bf 2).}
At previous lattice conferences, 
reports of QCD + QED calculations using staggered lattices
have been reported by MILC%
~\cite{Basak:2008na}.
These computations use 
$2 +1$--flavor 
$\texttt{asqtad}$ 
configurations with non-compact photon fields in the quenched QED approximation. 
The photon fields have been fixed to Coulomb gauge, 
with the following zero modes deleted:
$\vec{A} (p_\mu = 0) \equiv 0$
and 
$A_4(\vec{p} = 0, x_4) \equiv 0$. 
To improve precision, 
multiple values of the electric charge were used, 
$e = \pm |e_{phys}|$,
$\pm 2 |e_{phys}|$. 
Pion masses range from 
$270$--$520 \, \texttt{MeV}$, 
on lattices with volumes 
$(2.4 \, \texttt{fm})^3$, $(2.52 \, \texttt{fm})^3$ and $(3.0 \, \texttt{fm})^3$, 
and lattice spacings, 
$0.12 \, \texttt{fm}$, $0.09 \, \texttt{fm}$ and $0.125 \, \texttt{fm}$, 
respectively. 
Control over systematics was sought using rooted staggered chiral perturbation theory with QED.

{\bf 3).}
At the previous lattice conference, 
reports of QCD + QED simulations were given by the BMW collaboration%
~\cite{Portelli:2010yn}, 
with further updates provided at this conference%
~\cite{Portelli}.
These computations use an extensive ensemble of 
$2+1$--flavor BMW lattices with quenched non-compact QED. 
The photon field is gauged fixed to Coulomb gauge with zero modes deleted; 
the physical value of the electric charge is used. 
Pion masses range from 
$135$ 
to 
$422 \, \texttt{MeV}$, 
for which an interpolation to the physical point can be made, 
and volumes range from
$(2.0 \, \texttt{fm})^3$ 
to 
$(5.5 \, \texttt{fm})^3$. 
For these computations, 
a single lattice spacing has been utilized so far, 
$a = 0.12 \, \texttt{fm}$. 
Because of chiral symmetry breaking present for Wilson quarks, 
isospin asymmetric quark masses are radiatively generated at 
$\mathcal{O} ( \alpha_{f.s.} a^{-1} )$. 
This additive mass renormalization is removed by retuning to the isospin symmetric point. 
Finite volume effects appear to be well described by a power-law dependence on the volume, 
$\propto L^{-2}$, 
rather than predicted from chiral perturbation theory + QED. 
Partially quenched chiral perturbation theory is employed to estimate the effect of unquenching QED. 
While corrections to individual hadron masses are estimated at the sub-percent level, 
corrections to meson mass splittings are estimated at 
$\sim 5 \%$.

{\bf At this conference:}
Results of QCD + QED simulations were discussed by three groups. 
We heard of a preliminary study of the electromagnetic corrections to pseudoscalar decay constants
of light-light mesons and heavy-light mesons%
~\cite{Glaessle}.
Unique to this study is the use of compact photon fields. 
In another talk, 
we heard updates from the BMW collaboration%
~\cite{Portelli}, 
which we have included in our summary above.

We also heard preliminary reports from a comprehensive study by PACS-CS%
~\cite{Ukita}. 
This study is the first to report success at re-weighting configurations to include dynamical QED. 
Re-weighting configurations to include QED effects appears to be suggested first in~\cite{Duncan:2004ys}. 
Statistical noise in the re-weighting factor is shown to be a considerable hinderance---even
for a test case of re-weighting from 
$e = 0$ 
to 
$e = \frac{e_{phys} }{100}$. 
On a QCD ensemble of size 
$32^3 \times 64$, 
the noise was overcome by generating QED configurations on a lattice of twice finer resolution, 
$64^3 \times 128$, 
and subsequently averaging over all paths in a unit cell to form QED gauge links on the 
$32^3 \times 64$ 
lattice. 
The masses of the 
$\pi^+$, 
$K^+$, 
$K^0$, 
and 
$\Omega^-$
are used to re-weight PACS-CS configurations to 
$1+1+1$--flavor 
QCD + QED configurations. 
The study neglects running of the QED coupling, 
and QED renormalization of the quark masses;
however, 
these are tiny effects.

\section{Outlook}

There have been numerous developments in the study of lattice QCD with classical and quantum electrodynamics.
By the next lattice conference, 
we might see: 
further studies of hadron correlation functions in weak, uniform external fields; 
detailed exploration of the QCD phase diagram in strong magnetic fields; 
computations from multiple groups of the isospin splittings of mesons and baryons; 
and precision determination of the light quark masses accounting for 
strong and electromagnetic isospin breaking.

\begin{acknowledgments}
Work supported in part by the 
U.S.~Dept.~of Energy, 
Office of Nuclear Physics,
under
Grant No.~DE-FG02-94ER40818, 
through a joint CCNY--RBRC fellowship, 
and by the Research Foundation of the CUNY.
I am grateful to my collaborators, W.~Detmold and A.~Walker-Loud, for fruitful collaboration, 
and to T.~Izubuchi for useful discussions. 
\end{acknowledgments}




\end{document}